**Wigner polarons probe the dynamics of a Wigner crystal in a monolayer semiconductor**

Lifu Zhang[1], Liuxin Gu[1], Haydn S. Adlong[2,3], Arthur Christianen[2,3], Eugen Dizer[4], Ruihao Ni[1], Rundong Ma[1], Suji Park[5], Houk Jang[5], Takashi Taniguchi[6], Kenji Watanabe[7], Ilya Esterlis[8], Richard Schmidt[4], Atac Imamoglu[2], You Zhou[1†]

[1]Department of Materials Science and Engineering, University of Maryland, College Park, MD 20742, USA
[2]Institute for Quantum Electronics, ETH Zürich, Zürich, Switzerland
[3]Institute for Theoretical Physics, ETH Zürich, Zürich, Switzerland
[4]Institut für Theoretische Physik, Universität Heidelberg, Heidelberg, Germany
[5]Center for Functional Nanomaterials, Brookhaven National Laboratory, Upton, NY 11973, USA
[6]Research Center for Electronic and Optical Materials, National Institute for Materials Science, 1-1 Namiki, Tsukuba 305-0044, Japan
[7]Research Center for Materials Nanoarchitectonics, National Institute for Materials Science, 1-1 Namiki, Tsukuba 305-0044, Japan
[8]Department of Physics, University of Wisconsin-Madison, USA

†To whom correspondence should be addressed: youzhou@umd.edu

**Wigner crystals—lattices made purely of electrons—provide a platform for studying correlation-driven quantum phase transitions. Despite extensive research, accessing the internal dynamics of Wigner crystals has remained challenging, with most experiments probing only static order or collective motion. Here, we demonstrate optical probing and manipulation of zero-field Wigner crystals and elucidate their static and dynamic properties in the frequency domain. We observe optical resonances that we identify as Wigner polarons—quasiparticles formed when the electron lattice is locally distorted by exciton–Wigner crystal coupling. We further achieve all-optical control of spins in the Wigner crystal, directly probing valley-dependent Wigner polaron scattering well above the magnetic ordering temperature and in the absence of any external magnetic field. Finally, we show optical melting of the Wigner crystal and observe different responses of the umklapp (static) and Wigner polaron (dynamic) resonances to optical excitation. Our results provide an avenue for understanding electron dynamics and achieving ultrafast optical control of interaction-driven quantum phase transitions in strongly correlated electron systems.**

Elucidating the dynamics of correlated-electron systems is key to understanding these many-body quantum systems because it reveals the underlying interactions, collective modes, and phase competition, and may enable the control or creation of nonequilibrium phases for device applications[1-4]. Yet experimentally accessing such dynamics can be extremely challenging. A prime example is the Wigner crystals (WCs), where most experiments to date probe only static order or collective motion (e.g., pinning modes)[5-9], while their internal dynamics remain largely unexplored. This is partly because many dynamical signatures are not directly accessible with standard techniques, such as DC transport. Moreover, WCs are often fragile, with dynamical energy scales that are rather small, and therefore require ultralow temperatures and exceptionally clean samples to observe.

Understanding how these exotic crystals form and melt can provide fundamental insights into the competition between distinct quantum phases in diverse correlated systems[8, 10-15]. Historically, experimental studies of WCs in solid-state materials have predominantly required strong magnetic fields[8, 9, 16-20] to suppress electron kinetic energy, leaving Coulomb interaction as the dominant energy scale. The realization and investigation of zero-field quantum Wigner crystals have remained elusive[21, 22] until recent breakthroughs in high-quality two-dimensional (2D) transition metal dichalcogenide (TMD) monolayer and bilayer systems[5, 6, 23, 24]. These studies typically employed optical spectroscopy[5, 6] or scanning probe microscopy[23] to probe charge compressibility and reveal electron ordering within the WC.

Aside from recent THz measurements of WC pinning mode[7], the dynamics of Wigner crystals—and crucially the interplay between the crystallized electrons and optically excited excitons—remain largely unexplored. For instance, excitons can strongly couple to electrons, forming polarons[25-27] that appear robust even when electrons are organized into lattices[5, 6, 24, 25]. However, how excitons interact with Wigner crystals in this non-perturbative regime of strong electron-electron and strong electron-exciton interactions is not well understood. To date, it is unclear how exciton–electron interactions influence the spin order and overall stability of the Wigner crystal phase. Addressing these fundamental questions is critical to advancing our understanding of correlated phases and leveraging them for future quantum electronic and optoelectronic technologies[28-30].

## Results

### Probing dynamic properties of Wigner crystals

We first demonstrate $WSe_2$ monolayers as a materials platform hosting WCs and optically probe the dynamic properties of WCs. We fabricate a device made of a monolayer $WSe_2$ encapsulated inside hBN with graphite as a gate electrode (**Fig. 1a**). **Figure 1b** shows the reflectance contrast $R_C$ ($R_C = \Delta R/R_0$) in monolayer $WSe_2$ at 5 K in the low-density electron-doped regime, where $R$ is the reflectance on $WSe_2$ and $R_0$ is the background reflection from a nearby region without $WSe_2$ (see Methods). In the charge-neutral regime at $V < 0$, we observe a sharp 1s exciton resonance along with Rydberg states, including 2s and 3s excitons[31]. When the sample is electron-doped ($V > 0$), we observe a repulsive polaron (RP) branch whose energy blueshifts with increasing density, accompanied by two lower-energy attractive polarons (APs), triplet ($T^-$) and singlet ($S^-$), corresponding to inter- and intra-valley charged excitons[32].

To detect the formation of Wigner crystals, we focus on the umklapp scattering of excitons by the Wigner crystal. In the limit of weak exciton-carrier interactions, the WC induces a periodic potential for excitons that folds high-momentum dark excitons into the light cone and produces an additional bright exciton resonance. **Figure 1c** shows the voltage derivative of the reflectance contrast, $\mathrm{d}R_C/\mathrm{d}V$, where we observe the clear secondary resonance at higher energy above the RP peak. This feature exhibits a pronounced blueshift with increasing doping density and becomes indiscernible at carrier densities above $n_c > 7\times10^{11}$ cm$^{-2}$ (see Methods for the density calibration). We attribute this higher-energy resonance to the umklapp resonance of the RP, indicating the formation of a WC in $WSe_2$ below $n_c$ and its melting above $n_c$, similar to the case of $MoSe_2$[5]. We have measured multiple devices and several beam-spot locations, and generally observe a similar critical melting density when umklapp is observed (**Fig. S1**).

In addition to the umklapp resonance of the RP, we observe features originating from WCs above the APs. Specifically, two higher-energy resonances appear above the $T^-$ and $S^-$ states, which we term Wigner polarons and denote as $WP_T$ and $WP_S$, respectively. While the energies of the $T^-$ and $S^-$ attractive polarons remain nearly unchanged, $WP_T$ and $WP_S$ blueshift with increasing carrier density.

Whereas the umklapp feature associated with the repulsive polaron probes the static properties of the Wigner crystal, Wigner polarons are quasiparticles that arise from the dynamical properties of the WC, in particular its vibrational modes (**Fig. 2**). In the WC, the APs arise from collective trion formation across the Wigner-crystal sites analogously to Fermi polaron formation in continuous systems[26, 33]. This trion-creation process can, however, also locally excite the lattice. The WP feature is therefore naturally explained as an AP dressed by additional phonon excitations of the Wigner crystal (Ref [34] and Supplementary Discussion 1). Consequently, the WP-AP splitting is a direct measure of the characteristic phonon frequency of the Wigner crystal. Since the singlet and triplet trions have the same mass and charge, their coupling to the Wigner crystal lattice is identical. As a result, their corresponding WPs exhibit the same splitting relative to their respective AP branches. By contrast, there exists only one repulsive polaron and one umklapp peak because these features originate from exciton-electron scattering states, rather than bound states, making them relatively insensitive to the electronic spin[34]. We note that recent THz spectroscopy measurements have probed the collective pinning modes of the WC[7], whereas our optical studies provide a direct means to access the internal dynamics of the WC and associated quasiparticles.

**Quantitative analysis of density dependence**

Next, we quantitatively analyze the doping dependence of the umklapp resonance and Wigner polarons. The RP and umklapp energies are extracted by fitting reflectance-contrast spectra and their voltage/energy derivatives, which yield quantitatively consistent results (see Methods and **Figs. S2–S4** for density calibration and peak-energy extraction). The extracted RP-umklapp splitting $\Delta E_U$ linearly increases with $n$ (**Fig. 2b**), as theoretically predicted by $\Delta E_U = h^2 n/(\sqrt{3} m_X)$ where $h$ is the Planck constant and $m_X$ is the exciton mass[5]. Fitting our $\Delta E_U - n$ data to this relation yields an exciton mass of $m_X \sim (0.85 \pm 0.8)\, m_0$ ($m_0$ is electron mass), consistent with lower carrier masses in $WSe_2$[35] than $MoSe_2$ and in agreement with prior measurements ($m_X \sim 0.81 m_0$) [36, 37]. The uncertainty is dominated by the ~ 10% uncertainty in the carrier density (see Methods). From the disappearance of the umklapp resonance, we estimate that the WC melts when $r_s$, the Coulomb-to-Fermi energy ratio, approaches close to 12, using an effective electron mass of $0.4m_0$ and a dielectric constant of 4.5. This measured $r_s$ is smaller than the theoretical prediction in the clean limit[38, 39], suggesting an enhanced stability of WC in $WSe_2$. The enhancement could arise from charge disorder[7, 40-43], isovalent disorder[44], and Berry curvature effects[45]. This enhancement is even stronger than previously observed in $MoSe_2$[5], possibly reflecting differences in the disorder landscape and/or stronger Berry curvature in $WSe_2$. This enhancement and discrepancy merit further investigation.

For the Wigner polarons, the theoretical WP-AP energy splitting based on the phonon model, $\Delta E_{WP}$, scales as $n^{3/4}$ (see Supplementary Discussion 1), which reflects the density-dependence of the characteristic WC phonon energies (e.g., the WC Debye frequency). We find that the experimental peak positions can be well described by $\Delta E_{WP} = an^{3/4}$ (**Fig. 2d**), with only one free parameter $a$, which characterizes the stiffness of the Wigner crystal. The experimentally extracted $a$ is around 1.5 times the value predicted by a minimal phonon model (see Supplementary Discussion 1), indicating good quantitative agreement. The slightly larger experimental value implies a stiffer Wigner crystal than the model, potentially related to the crystal's unusual robustness and the effects of disorder. While the $n^{3/4}$ scaling is consistent with the experimentally measured WP-AP splitting, the power law cannot be unambiguously extracted from the data, largely because WP energies cannot be reliably determined at low density. The weak WP signal is likely due to the intrinsically small WP spectral weight in this regime, compounded by disorder, which further reduces visibility through inhomogeneous broadening. For example, a linear fit is also possible but would require a finite intercept at the onset of doping (**Fig. S5**). Elucidating the exact density dependence of $\Delta E_{WP}$, together with its linewidth and oscillator strength, will shed further light on the WC phonons and exciton-phonon coupling, and will likely require resolving WP peaks in the lower density regime, which merits future studies.

While the present work focuses on electron doping, we also observe an umklapp resonance under hole doping, indicative of hole WC formation, similar to a recent report[46]. Moreover, a higher-energy resonance above AP can be observed on the hole side, indicative of WP on the hole side (**Fig. S6**). The WP-AP splitting on the hole side shows a similar density dependence to that on the electron side.

**Optical control of WC spin and Wigner polarons**

At sufficiently low temperature, the WC should magnetically order[11, 12, 15, 39, 47, 48], which would have a profound impact on the exciton-WC interactions. However, the expected magnetic ordering temperatures of WC are extremely low[11, 12, 15], making it difficult to directly probe exciton–WC interactions in a spin-ordered phase without an external magnetic field.

Here, we investigate the optical control of WC spin polarization and explore how excitons interact with spin-ordered electron arrays to form Wigner polarons (**Fig. 3a**). To do this, we excite the system using circularly polarized light and measure the reflectance spectra under different polarization configurations. Previous studies have shown that circular pumping can induce strong spin polarization in electron-doped $WSe_2$[49], likely due to distinct scattering dynamics between inter-valley and intra-valley processes, although the exact mechanism remains under investigation.

As shown in **Fig. 3b**, under co-circular polarization, the reflectance is dominated by the triplet feature, while under cross-circular polarization, the singlet feature becomes dominant (**Fig. 3c**).

This contrast indicates that electrons become strongly valley(spin)-polarized in the −K valley when excitons are optically excited in the K valley. With electrons polarized in −K, excitons in K (-K) can scatter only via intervalley (intravalley) processes, yielding a single WC-polaron peak, $WP_T$ ($WP_S$), under co-polarization (cross-polarization). Therefore, the observation confirms the one-to-one correspondence between the two Wigner polarons and the two AP branches, as predicted by our theory[34].

Furthermore, consistent with our theoretical model (Ref[34] and Supplementary Discussion 1), this valley- and spin-dependent scattering behavior differs markedly between the AP and RP branches. **Figure 3d** presents line cuts of the $dR_C/dV$ spectra, comparing spin-polarized and unpolarized WCs at a fixed doping level. While the AP and Wigner crystal polarons exhibit distinct changes in scattering under spin-polarized conditions (stronger $T^-/WP_T$ under co-polarization and stronger $S^-/WP_S$ under cross-polarization), the umklapp scattering of the RP branch remains largely unaffected (**Fig. 3d**).

**Optical melting of WC**

Finally, we demonstrate the optical modification of the Wigner crystals and Wigner polarons by tuning the exciton density. To achieve this, we excite the system using a pulsed supercontinuum laser with a pulse duration of about 200 ps and a 40 MHz repetition rate. The laser energy is spectrally filtered to lie below the free-carrier bandgap of $WSe_2$ (about 1.9 eV[31]), while covering the exciton resonance. This approach allows us to selectively excite excitons without generating a substantial number of free carriers, while simultaneously measuring their reflectance spectrum.

**Figures 4a** and **4b** show the $dR_C/dV$ maps of monolayer $WSe_2$ under different pump powers. While optical excitation has little effect on the RP, AP, and WP branches, it strongly influences the umklapp resonance. In particular, with increasing pump power, the umklapp resonance of the RP state becomes weaker (**Figs. 4** and **S7**), and nearly vanishes at a moderate power of about 4000 nW, suggesting optical melting of the Wigner crystal. By contrast, the Wigner polarons experience remain robust but experience linewidth broadening and a slight redshift at this pumping power (**Fig. S11**).

The mechanism underlying this optical melting and the distinct behaviors of the umklapp and WP features remain to be fully clarified, but several scenarios are plausible. Laser-induced heating is an obvious possibility. However, if the lattice temperature approached the WC melting point, the 1s exciton would be expected to redshift by about 2 meV (see Supplementary Discussion 2 and **Figs. S8**, **S9**). Instead, the 1s exciton energy remains essentially unchanged across pump power (**Fig. S10**). This indicates negligible lattice heating, though we note that the electronic temperature may exceed the lattice one, which merits further study. Second, optical doping could weaken the WC. However, we do not observe appreciable shifts in the RP and AP branch energies, arguing against a simple carrier–density change as the dominant mechanism.

Another possibility is that excitons themselves destabilize the Wigner crystal, through exciton-electron scattering. The observed WP redshift and linewidth broadening are compatible with softening of WC phonon modes, possibly due to enhanced exciton screening of electron correlation (**Fig. S11**). The WP resonances appear to persist to higher optical powers than the umklapp resonance, potentially because the umklapp feature is more sensitive to the reduction of static charge-density correlations at the relevant wavevector. On the other hand, the WP resonances are more sensitive to local short-range correlations and therefore can persist even in a partially melted regime. To access the magnitude of this potential effect, we examine how the critical electron density for Wigner crystallization, $n^*$, evolves with exciton density (**Fig. 4c**). We estimate that the bright exciton density reaches about $10^9\ cm^{-2}$ near the critical optical pump threshold (see Methods and Ref[50]), far below the WC critical density $n_c$, implying that bright excitons alone are unlikely to melt the WC. However, in monolayer $WSe_2$, strong spin–orbit coupling yields lower-energy, long-lived dark excitons that are nominally spin-forbidden[32, 51, 52]. Given their longer lifetimes[52, 53], we estimate a dark exciton population of approximately $10^{11}\ cm^{-2}$. While our estimate for dark exciton density is still lower than $n_c$, possible exciton-mediated softening cannot be ruled out. Further studies will be needed to quantitatively compare the pump-power dependence of the umklapp and WP spectral weights. Irrespective of the detailed mechanism, however, the different response of the umklapp and WP resonances to optical excitation suggests that further theoretical and experimental work is needed to develop a complete understanding of the signatures of Wigner crystals.

## Conclusion

Optical probing and manipulation of zero-field WCs in monolayer $WSe_2$ opens exciting avenues for studying their dynamics and quantum optoelectronics. Our AP spectroscopy directly resolves Wigner polarons, providing critical insight into the internal phonon dynamics of WCs. These insights can help clarify the stability of correlated states—including WCs and generalized Wigner crystals—against thermal and quantum fluctuations, and the nature of the associated phase transitions. Optical excitation of such Wigner polarons could offer a direct handle on electronic vibrations and a route to driving interaction-induced quantum phase transitions. The demonstrated optical control of spin could enable optoelectronic and quantum devices that leverage the strong magnetic susceptibility near the quantum melting of WCs[43] for efficient spin manipulation. Further studies of exciton-assisted optical melting will elucidate the interplay of exciton–electron and electron–electron interactions, revealing new aspects of quantum criticality and enabling ultrafast optical control of quantum phase transitions beyond conventional density control. Finally, extending these techniques to electrostatically coupled bilayers—with their even richer phase diagram[54]—would enable studies of the dynamics of bilayer Wigner crystals[6], excitonic Wigner crystals[55], and excitonic insulators[56].

**Funding:**
This research is primarily supported by the U.S. Department of Energy, Office of Science, Office of Basic Energy Sciences Early Career Research Program under Award No. DE-SC-0022885. The fabrication of samples is supported by the National Science Foundation CAREER Award under Award No. DMR-2145712 and ARO W911NF2510066. This research used Quantum Material Press (QPress) of the Center for Functional Nanomaterials (CFN), which is a U.S. Department of Energy Office of Science User Facility, at Brookhaven National Laboratory under Contract No. DE-SC0012704. I.E. was supported by the National Science Foundation (NSF) through the University of Wisconsin Materials Research Science and Engineering Center Grant No. DMR-2309000. K.W. and T.T. acknowledge support from the CREST (JPMJCR24A5), JST and World Premier International Research Center Initiative (WPI), MEXT, Japan. E.D. and R.S. acknowledge support from the DFG (German Research Foundation) – ProjectID 273811115 – SFB 1225 ISOQUANT, and Germany's Excellence Strategy EXC 2181/1 - 390900948 (the Heidelberg STRUCTURES Excellence Cluster). A.C. is supported by an ETH Fellowship.

**Author contributions**
Y.Z. and L.Z. conceived the project. L.Z. fabricated the samples and performed the experiments. L.G., R.N., R.M., S.P., and H.J. assisted with sample fabrication. L.G. and R.M. helped with optical measurements. H.S.A., A.C. E.D., A.I., and R.S. contributed to the theoretical interpretation of the data. L.Z., I.E., and Y.Z. contributed to the data analysis. T.T. and K.W. provided hexagonal boron nitride samples. L.Z. and Y.Z. wrote the manuscript with extensive input from the other authors.

**Competing interests**
The authors declare no competing interests.

**Figure captions**

**Figure 1. Signature of WC in $WSe_2$. a**, A schematic of the device structure. **b**, Reflectance contrast ($R_C$) in the low-density electron-doped regime of monolayer $WSe_2$. The repulsive exciton polarons of 1s, 2s and 3s are indexed. $T^-$ and $S^-$ donate triplet and singlet, respectively. **c**, Voltage derivative of the reflectance contrast ($dR_C/dV$) in the low-density electron-doped regime. The weak, higher energy resonance $U_{RP}$ above the RP is due to umklapp scattering of the excitons off the WC. $WP_T$, $WP_S$ are corresponding Wigner polaron resonances of $T^-$ and $S^-$, respectively.

**Figure 2. Exciton umklapp and Wigner polarons in a Wigner crystal. a,** A schematic of repulsive polarons in Wigner crystals. **b**, Energy splitting $\Delta E_U$ for RP determined as a function of doping density. The solid line is a linear fit to the experimental data. Inset, umklapp scattering from the Wigner-crystal lattice ($k_W$) folds the exciton band, producing a zero-momentum resonance at a higher energy. **c,** A schematic of attractive polarons creating Wigner crystal polarons. **d,** Voltage derivative of the reflectance contrast ($dR_C/dV$), revealing Wigner-polaron resonances. Dashed lines show the corresponding theoretical density dependence $\Delta E_{WP} = an^{3/4}$, with a single free parameter $a$. Here the experimentally extracted $a = 4.16 \pm 0.11$ meV/ $(10^{11}\ \mathrm{cm}^{-2})^{3/4}$.

**Figure 3. Optical control of WC spin. a**, A schematic of optical control of WC spin polarization via circular pumping. **b**, The voltage derivative of the reflectance contrast ($dR_C/dV$) under co-circular polarization configuration ($\sigma$+ pump, $\sigma$+ probe). The reflectance is dominated by the triplet feature. **c**, The voltage derivative of the reflectance contrast ($dR_C/dV$) under cross-circular polarization configuration ($\sigma$+ pump, $\sigma$- probe). The reflectance is dominated by the singlet feature. **d**, Representative line cuts of the $dR_C/dV$ spectra, comparing spin-polarized (blue and orange lines) and unpolarized (yellow line) WCs at a fixed doping level. $T^-$ and $WP_T$ resonances are enhanced under co-polarization, while $S^-$ and $WP_S$ resonances are enhanced under cross-polarization, compared to the case of unpolarized WC. Note that **c** and **d** share the same color bar.

**Figure 4. Optical melting of WC. a, b**, The voltage derivative of the reflectance contrast ($dR_C/dV$) of monolayer $WSe_2$ in the low-density electron-doped regime under pulsed pumping of 160 nW (**a**) and 4000 nW (**b**). Note that **a** and **b** share the same color bar. **c**, The critical electron density for Wigner crystallization as a function of estimated injected exciton density under pulsed pumping. The bottom/top axis represents the injected exciton density estimated with pure bright/dark excitation, respectively. Error bars come from the chosen range of $dR_C/dV$ (-0.005 to -0.0065) to determine the cutoff carrier density. Inset, schematic of optical melting of WC driven by exciton screening of electron repulsion.

**Methods:**

**Device design and fabrication**

Graphite (HQ graphene Inc.), hBN flakes (provided by World Premier International Research Center Initiative (WPI), MEXT, Japan for hBN synthesis) were mechanically exfoliated from the bulk crystals onto the $SiO_2$ substrate. $WSe_2$ flakes are provided by the Quantum material press (QPress) facility in Center for Functional Nanomaterials (CFN) at Brookhaven national laboratory (BNL). The layer numbers of $WSe_2$ were estimated based on the color contrast under the optical microscopy, while the thickness of hBN dielectric layers (around 31 nm and 20 nm for top and bottom hBN layers respectively in Device 1, for example) was measured by AFM (Asylum Research Cypher). The heterostructure was fabricated in a transfer station (Everbeing Int'l Corp.), which uses PDMS (Polydimethylsiloxane) and PC (Polycarbonate) as a soft stamp and transfers all the flakes in a dry transfer method onto a silicon chip with 285 nm $SiO_2$ layer. The electrical contacts were patterned by electron-beam lithography after which we deposited 5 nm of Cr and 90 nm of Au by thermal evaporation.

**Optical spectroscopy and experimental setup**

The optical measurements were performed in our home-built confocal microscope with Attodry 4K cryostat (AttoCube 800). An apochromatic objective with a numerical aperture NA=0.82 is equipped in the chamber. The reflectance measurement is performed using a supercontinuum white laser (YSL Photonics Inc.) as the excitation source. The white laser has a pulse duration of ~200 ps with variable repetition. Normally an excitation power of 40 nW (40 MHz) is adopted. The spectra are collected by a Horiba iHR320 spectrometer using a 600 mm/line grating and a Synapse-Plus back-illuminated deep depletion CCD camera. The reflectance spectrum $R$ is normalized by dividing the reflected light intensity $R_0$ from the nearby bare hBN region. We define the reflectance contrast as $R_C \equiv \frac{\Delta R}{R_0} = \frac{R-R_0}{R_0}$ and the derivative of $R_C$ versus gate voltage as $R'_C(V_{g,n}) = [R_C(V_{g,n+2}) - R_C(V_{g,n})]/[V_{g,n+2} - V_{g,n}]$. Here $V_g$ denotes either the top-gate ($V_t$) or back-gate ($V_b$) voltage. In the experiments of optical control of WC spin, we excite the system with 635 nm circularly polarized light (4 μW) and measure the reflectance spectra under different polarization configurations. In the experiments of optical melting of WC, a pulsed white laser is utilized as the pumping source. We directly excite the system with a white laser of different averaged powers (40 to 8000 nW). Note that it has a pulse frequency of 40 MHz and is filtered to 650-800 nm wavelength range. In the experiments of thermal melting of WC, we use the Attocube thermal coupler ATC100 module which allows the sample temperature to be fine tunable in a wide range by controlling the heating power through resistors.

To better visualize weak features (e.g., WP) alongside stronger structures such as AP and RP, we modified the standard blue–white–red diverging colormap. In the conventional version, color varies linearly with the plotted value. Here, we apply a nonlinear intensity scaling by mapping values through a power-law transform with exponent $p < 1$, which increases color contrast near zero while compressing the extremes. This adjustment enhances variations near zero values without altering the underlying data values; only the color mapping is changed. We use $p$=0.6 for **Figs. 1-4**, **S1-2**, **S6-8**, and **S11**.

**Determination of energy of excitons and Umklapp resonances**

There is a reflection at each interface between two nearby layers in our device so the light reflected off the target monolayer will interfere with the rest reflected signals (including background). It sizably alters the line shape of the excitonic resonances observed in the spectra. To account for this effect, we describe each of these resonances using an effective dispersive Lorentzian spectral profile[57]: $R_X = \frac{A}{(E-E_X)^2+\gamma^2/4}[\gamma/2 cos\, \alpha_0 - (E - E_X) sin\, \alpha_0\ ] + C$, where $E$ denotes the photon energy, $A$, $E_X$ and $\gamma$ respectively, correspond to the amplitude, energy, and linewidth of the resonance $X$, $\alpha_0$ stands for interference-induced phase shift and $C$ represents a flat background.

To extract the energies of the excitonic resonance and Umklapp peak from a given reflectance contrast spectrum $R_C$, we first fit the spectral profile of the excitonic peak with the aforementioned dispersive Lorentzian formula. **Figure S3a** shows the result of such a fit performed for monolayer $WSe_2$ encapsulated in hBN at $V_t = 0.81\ V$. We then subtract the fitted line shape $R_X$ from the original data $R_C$ to obtain the Umklapp part. There are three methods we have adopted to extract the Umklapp peak. In the first method, we perform the derivative of $R_C - R_X$ with respect to $V$, as displayed in **Fig. S3b**. The spectrum can be divided into three ranges: $E < E_X - \gamma$, $E_X - \gamma < E < E_X + \gamma$ and $E > E_X + \gamma$. As we can see, there is a local maximum for the $E > E_X + \gamma$ part, which is linked to Umklapp and can be regarded as its peak position. For the second method to extract the Umklapp peak, we can do the derivative of $R_C - R_X$ with respect to $E$, as displayed in **Fig. S4b**. There is a local minimum for the $E > E_X + \gamma$ part, corresponding to an inflection point in the Umklapp region (highlighted by a dashed line in **Fig. S4a**). This feature is also linked to Umklapp and its voltage evolution can be obtained by repeating such procedure for each gate voltage, as displayed in **Fig. S4c**. Clearly, the inflection point shifts as gate voltage changes while another feature in the $E < E_X - \gamma$ range does not change much, further providing strong evidence for its link to Umklapp. In the last method, we directly fit the Umklapp peak from $R_C - R_X$ with a dispersive Lorentzian spectral profile, as shown in **Fig. S4d**. To avoid the spurious contribution from residuals of the excitonic fitting, the fitting range of the Umklapp resonance is truncated to the energies $E > E_X + \gamma$. **Figure S5** shows the resulting doping dependence of $\Delta E_U$, determined with all these three methods. As we can see, the results from different procedures basically agree with each other, confirming the solidity of our processing methods to extract energies of the exciton and Umklapp resonances.

**Doping control and estimation of doping density**

A single-gate scan is used to implement doping control in monolayer $WSe_2$. The doping density is determined by considering the heterostructure as a parallel capacitor. The top and bottom capacitance are given by $C_{Top(Bottom)} = \frac{\varepsilon_0 \varepsilon_{hBN}}{t_{hBN}}$, where $t_{hBN}$ is the thickness of top and bottom hBN dielectric layers, and they are extracted by atomic force microscope measurements. The total doping density in the system can be determined as $n = \frac{1}{e} \cdot (C_{Top} \cdot \Delta V_t + C_{Bottom} \cdot \Delta V_b)$ . $\Delta V_t$ and $\Delta V_b$ is the applied top and bottom gate voltage relative to the onset voltages of doping, respectively. The doping onset voltage for electron doping is determined by examining the reduction in the oscillation strength of RP and the shift in the 2s exciton energies, both of which are sensitive to doping levels (**Fig. S12**). We use $\varepsilon_{hBN} = 3.5$[58] in our case to estimate doping density. Recent experiments that calibrate carrier density from magnetic studies suggest a lower dielectric constant of $\varepsilon_{hBN} = 3.1$[59]. Therefor we estimate a density uncertainty in the density of about 10%. The electron density can also be estimated by comparing the measured RP blueshift with the $E_{RP}$-$n$ dependence of a sample with known density $n$, which yields qualitatively similar density estimation[60].

**Estimation of injected exciton density and Umklapp critical density**

In the CW pumping case, the injected exciton densities can be estimated by $n_X = P\alpha\Gamma/A\hbar\omega$, where $P$ is the diode laser pump power, $A$ is the pump beam size, $\hbar\omega$ is the photon energy, $\alpha$ is the sample's absorbance at the pump wavelength (10% for example), and $\Gamma$ is the lifetime of respective excitons. There are two types of excitons (bright intralayer exciton and dark intralayer exciton). A maximum (100% dark exciton) and a minimum (100% bright exciton) of injected exciton density can be calculated accordingly. In the pulsed pumping case, we convert the average pump power ($P_{Avg}$) into peak power $P = \frac{P_{Avg}}{f_R \cdot t_p}$, where $f_R$ is the repetition rate and $t_p$ is the pulse duration. For the estimation of Umklapp critical density under different conditions, a fixed differentiated reflectance contrast value (-0.005 to -0.0065 in optical melting experiments, for example) is deliberately chosen to determine the cutoff carrier density, beyond which WC is regarded to be melted into a liquid phase.

**Data availability**

Source data are provided with this paper.

**Methods-only references**

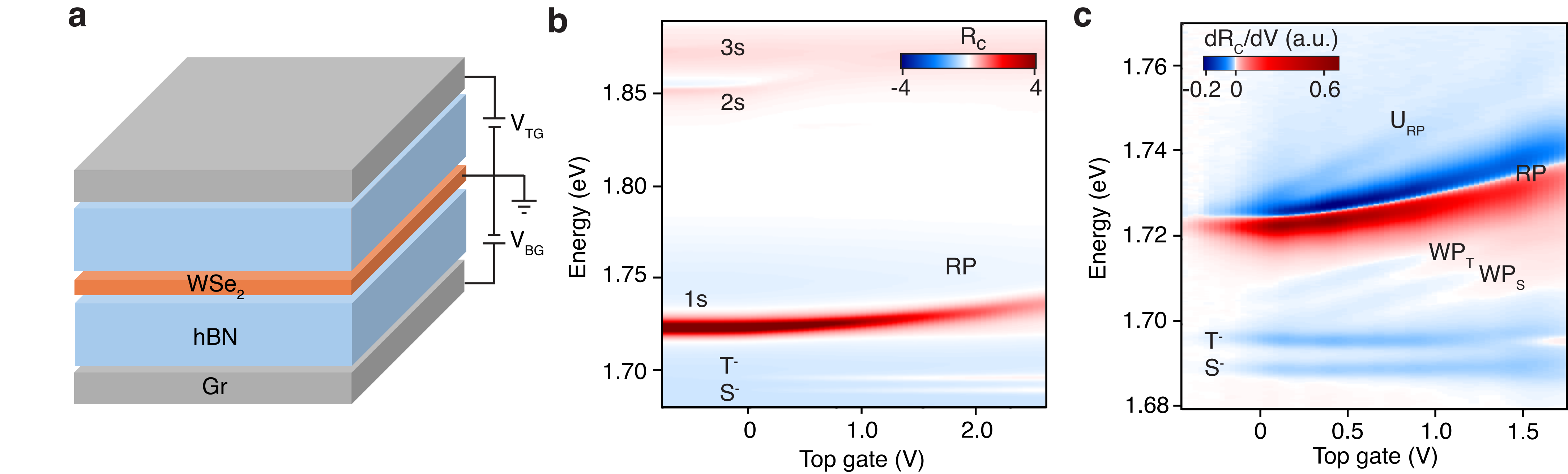

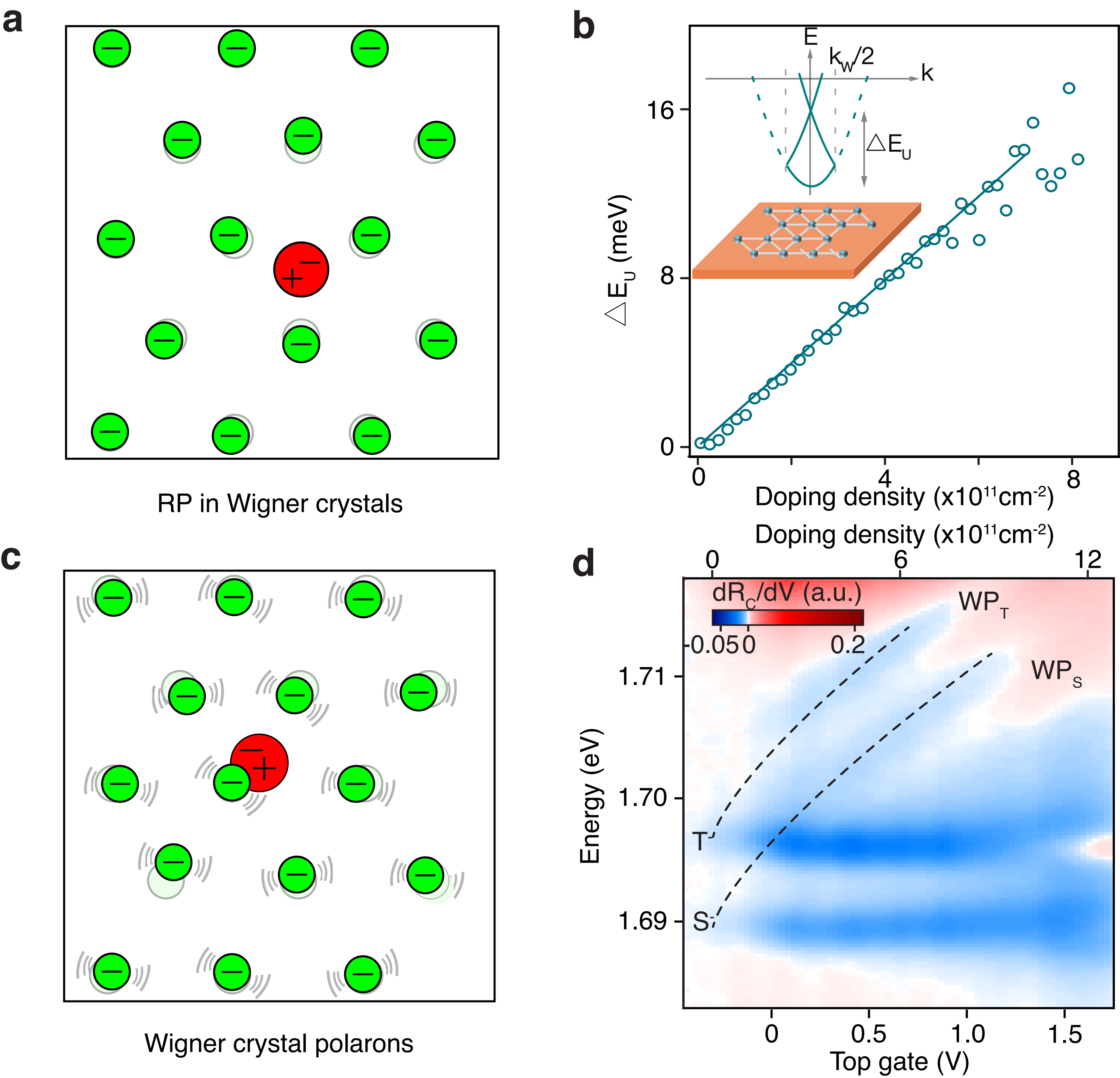

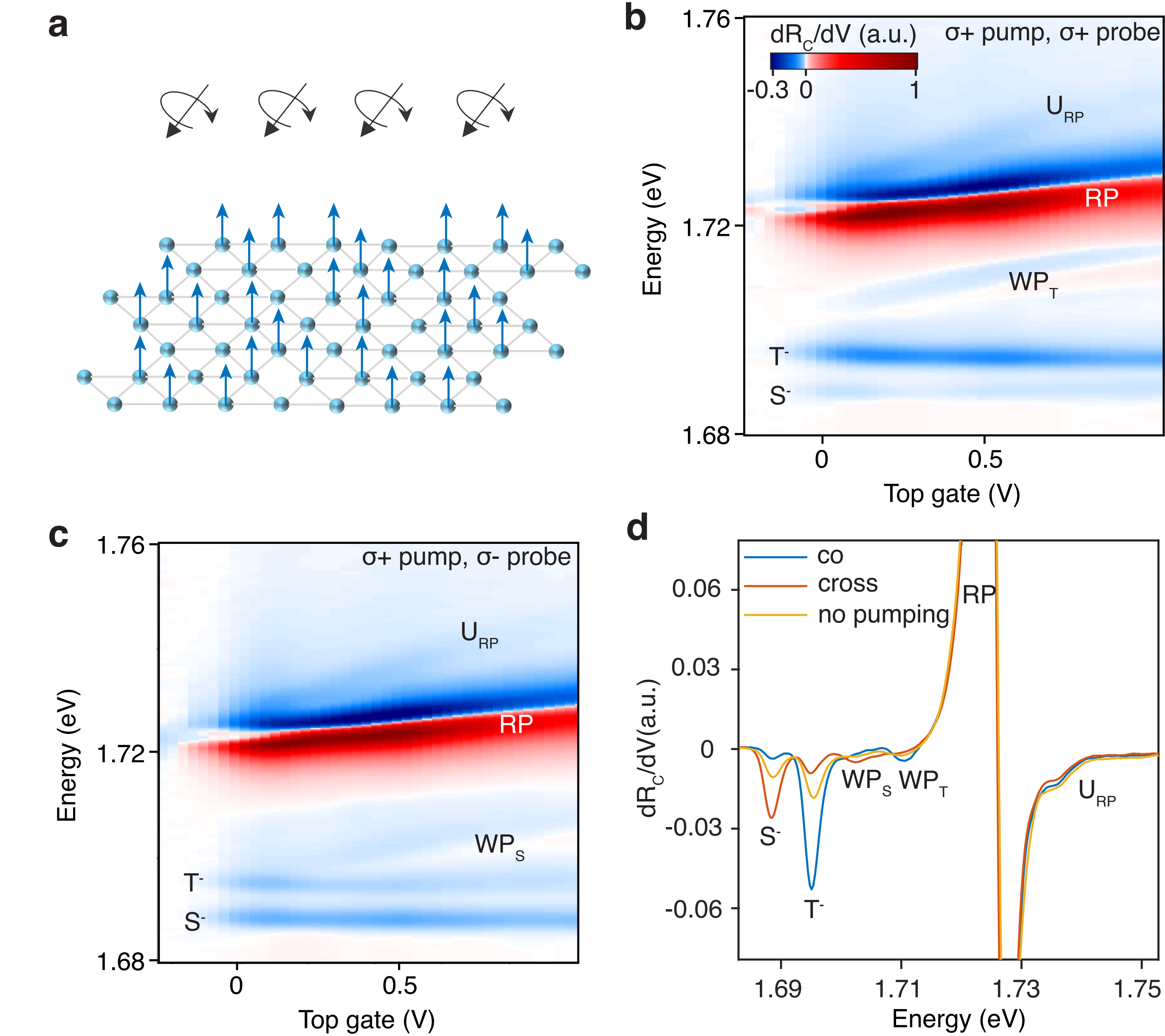

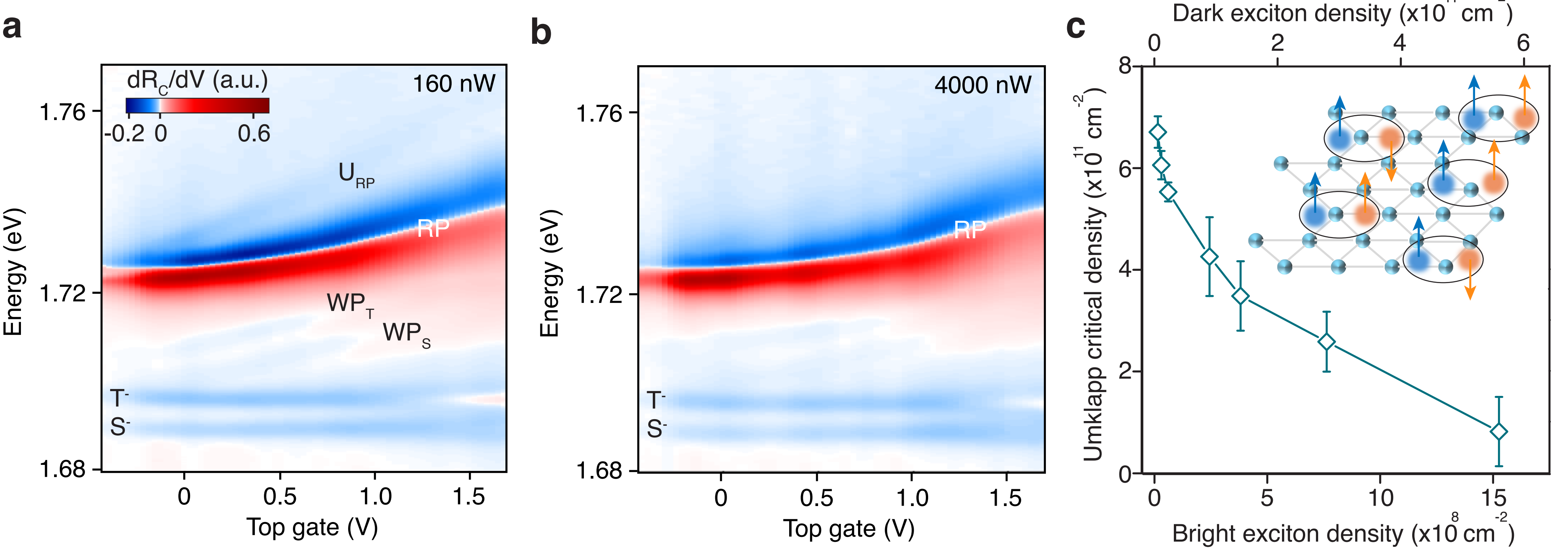